\documentclass{SciPost}

\hypersetup{
    colorlinks,
    linkcolor={red!50!black},
    citecolor={blue!50!black},
    urlcolor={blue!80!black}
}

\usepackage[bitstream-charter]{mathdesign}
\DeclareSymbolFont{usualmathcal}{OMS}{cmsy}{m}{n}
\DeclareSymbolFontAlphabet{\mathcal}{usualmathcal}

\fancypagestyle{SPstyle}{
\fancyhf{}
\lhead{\colorbox{scipostblue}{\bf \color{white} ~SciPost Physics}}
\rhead{{\bf \color{scipostdeepblue} ~Submission }}

\fancyfoot[C]{\textbf{\thepage}}
}
\usepackage{comment}
\begin{document}

\pagestyle{SPstyle}

\begin{center}{\Large \textbf{\color{scipostdeepblue}{Classical representation of the dynamics of quantum spin chains \\
}}}\end{center}

\begin{center}\textbf{
Tony Jin\textsuperscript
}\end{center}

\begin{center}
Universit\'e C\^ote d'Azur, CNRS, Centrale Med, Institut de Physique
de Nice, 06200 Nice, France.
\\[\baselineskip]
\href{mailto:email1}{\small tony.jin@univ-cotedazur.fr}\,
\end{center}

\section*{\color{scipostdeepblue}{Abstract}}
\textbf{\boldmath{%
Since the advent of quantum mechanics, classical probability interpretations
have faced significant challenges. A notable issue arises with the
emergence of negative probabilities when attempting to define the
joint probability of non-commutative observables. In this work, we
propose a resolution to this dilemma by introducing an exact representation
of the dynamics of quantum spin chains using classical continuous-time
Markov chains (CTMCs). These CTMCs effectively model the creation,
annihilation, and propagation of pairs of classical particles and
antiparticles. The quantum dynamics then emerges by averaging over
various realizations of this classical process. 
}}

\vspace{\baselineskip}

\noindent\textcolor{white!90!black}{%
\fbox{\parbox{0.975\linewidth}{%
\textcolor{white!40!black}{\begin{tabular}{lr}%
  \begin{minipage}{0.6\textwidth}%
    {\small Copyright attribution to authors. \newline
    This work is a submission to SciPost Physics. \newline
    License information to appear upon publication. \newline
    Publication information to appear upon publication.}
  \end{minipage} & \begin{minipage}{0.4\textwidth}
    {\small Received Date \newline Accepted Date \newline Published Date}%
  \end{minipage}
\end{tabular}}
}}
}


\vspace{10pt}
\noindent\rule{\textwidth}{1pt}
\tableofcontents
\noindent\rule{\textwidth}{1pt}
\vspace{10pt}


\section{Introduction}

The challenge of reconciling quantum mechanics (QM) with classical
stochastic processes has been a fundamental issue since the inception
of quantum theory. A critical aspect of this discussion involves the
concept of negative probabilities \cite{DiractNegativeproba,feynman1984negative}
which arise when considering joint distribution of non-commuting observables
\cite{AthurFine_hiddenvariables_jointproba,Fine_1982_long}, such
as e.g. the momentum-space distribution of a single particle \cite{WignerWignerfunction}.
The consensus established then, which continues to prevail, is that
negative probabilities lack intrinsic physical meaning and should
only be viewed as useful tools for facilitating intermediate calculations.

In this work, we propose a novel approach where the quantum dynamics
generated by Schr\"odinger evolution will be entirely interpreted as
a classical continuous-time Markov chains (CTMCs). We will focus on
quantum spin chains, where the lack of commutativity of Pauli operators
along different axis naturally gives rise to negative probability.
To get rid of the latter, we consider equivalent processes where the
probabilities are positive but the \emph{transition rates }become
negative. The advantage of this approach lies in its ability to leverage
a recent methodological advancement by V\"{o}llering \cite{Negativerates_Vollering},
which systematically maps CTMCs with negative transition rates to
equivalent CTMCs with entirely positive transition rates at the cost
of \emph{doubling} the configuration space. The expanded space can
be interpreted as introducing classical \textquotedbl antiparticles\textquotedbl ,
providing a clear physical analogy. The quantum dynamics of the system
emerges through the statistical averaging over realizations of the
classical stochastic process, establishing a direct connection between
classical and quantum descriptions.

We begin by explaining the emergence of negative probabilities in
the context of quantum spin chain dynamics. Next, we explain V\"ollering's
procedure for mapping a Markov process with negative transition rates
to a positive \cite{Negativerates_Vollering}. Finally we demonstrate
how this procedure applies to spin chains, starting with the simplest
case of a spin-$\frac{1}{2}$ rotation. We conclude with a discussion
of potential future directions.

\begin{figure}
\centering{}\includegraphics[width=0.9\textwidth]{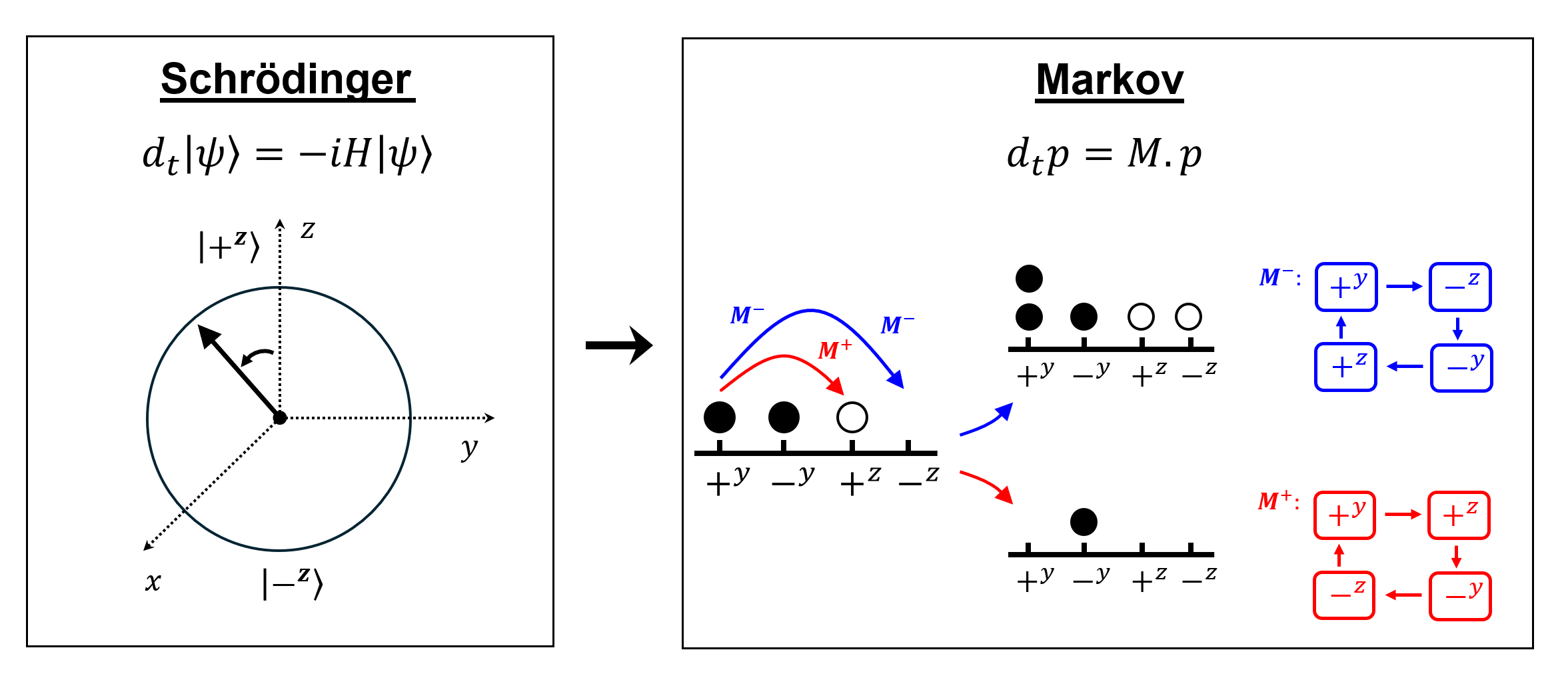}\caption{Our formalism establishes an exact correspondence between the dynamics
of spin chains and continuous-time Markov chains (CTMCs). In this
figure, we illustrate the mapping of the rotation of a spin-$\frac{1}{2}$
around the $x$ axis of the Bloch sphere to a four-states classical
CTMC featuring particles (black) and antiparticles (white) that annihilate
upon interaction. These particles(antiparticles) move according to
the rules fixed by the transition matrices $M^{\pm}$ (\ref{eq:Mpm}).
$M^{+}$ moves the particles(antiparticles) from a given configuration
to another while $M^{-}$ converts a particle(antiparticle) into its
opposite while simultaneously creating two more particles(antiparticles)
on the original site. Another possible interpretation
is that $M^{-}$ is really making a transition from the \emph{target
} to the original site. If the target is empty, the
price to pay is to create a particle with opposite sign. The Markov
transition rules for the example discussed in the main text with $H=\frac{\sigma^{x}}{2}$
are shown on the right. \label{fig:Examples-of-transitions}}
\end{figure}

\section{Negative Markov chains in quantum spin chains}

Throughout this work we will consider spin-$\frac{1}{2}$ chains of
$N$ sites living on an Hilbert space ${\cal H}=\mathbb{C}^{2\otimes N}$.
Let $H$ be any linear combination of Pauli operators strings $\sigma_{j_{1}}^{\alpha_{1}}\cdots\sigma_{j_{n}}^{\alpha_{n}}$
with $n\in\left[1,N\right]$, $\alpha_{j}\in\left\{ x,y,z\right\} $
and $\sigma^{\alpha}$ are the usual Pauli operators. Let $\rho_{t}$
be the density matrix of the system at time $t$. Our goal is to provide
a general description of the dynamics generated through the Schr\"odinger
evolution 
\begin{equation}
d_{t}\rho_{t}=-i\left[H,\rho_{t}\right]\label{eq:Schrodinger}
\end{equation}
in terms of a \emph{classical }CTMC\emph{.} We define the classical
configuration space ${\cal C}$ as the set composed of the $6^{N}$
elements 
\begin{equation}
{\cal C}=\times_{j=1}^{N}\left(\left\{ +,-\right\} \times\left\{ x,y,z\right\} \right)_{j}\label{eq:classicalconfigurationspace}
\end{equation}
where $\left\{ +,-\right\} $ denotes the orientation of the spin
along a given axis. We call a \emph{classical configuration} $C$
an element of ${\cal C}$ \footnote{Alternatively, we could have defined the joint probability on the
\emph{tensor product }space $\left\{ +,-\right\} _{x}\otimes\left\{ +,-\right\} _{y}\otimes\left\{ +,-\right\} _{z}$
leading to $8$ states on a single site instead of $6$. For a given
Hilbert space basis element $\left|\pm^{x/y/z}\right\rangle $ the
associated probability $p_{C}$ would then be seen as a \emph{reduced
}probability. This introduces the additional complexity of determining
the joint probability from the reduced ones (which is in general not
unique and non-positive) and it is in order to avoid this that we
work with the direct product. In future works, it may nevertheless
be advantageous to exploit the local tensor product structure.}. To each $C$, we can naturally associate the Hilbert space vector
$\left|C\right\rangle $. Note that $\left\{ \left|C\right\rangle \right\} _{C\in{\cal C}}$
is an overcomplete basis. Let $\mathbb{P}_{C}:=\left|C\right\rangle \left\langle C\right|$,
$\sum_{C}\mathbb{P}_{C}=\mathbb{I}$. From there, we define the probabilities
\begin{equation}
p_{C}(t)=\frac{1}{m^{N}}{\rm tr}\left(\rho_{t}\mathbb{P}_{C}\right).
\end{equation}
Because of the Hermiticity of $\rho_{t}$, $p_{C}\geq0$ and $m$
is chosen so that $\sum_{C\in{\cal C}}p_{C}=1$ so that $p_{C}$ is
a well-defined probability distribution. For the configuration space
(\ref{eq:classicalconfigurationspace}), $m=3$. %
The Heisenberg time evolution of the projection operator $\mathbb{P}_{C}$
is given by
\begin{equation}
d_{t}\mathbb{P}_{C}=i\left[H,\mathbb{P}_{C}\right].\label{eq:Heisenberg}
\end{equation}
The commutator $i\left[H,\mathbb{P}_{C}\right]$ can always
be decomposed as a linear combination of 
$\mathbb{P}_{C'}$, 
\begin{equation}
i\left[H,\mathbb{P}_{C}\right]=:\sum_{C'}M_{CC'}\mathbb{P}_{C'}
\end{equation}
where $M$ is an $6^{N}\times6^{N}$ matrix with \emph{real }entries. Note that, because the quantum basis is \emph{overcomplete}, this decomposition is not unique. In all the examples we consider later, we will explicitly fix this gauge degree of freedom. Taking the trace over $\rho_{t}$ and making use of the conservation
of the total probability, (\ref{eq:Heisenberg}) can be, without loss
of generality, written as 
\begin{equation}
d_{t}p_{C}(t)=\sum_{C'\neq C}\left(M_{CC'}p_{C'}-M_{C'C}p_{C}\right).\label{eq:evolutionp_C}
\end{equation}
This is almost the form of a CTMC on a discrete configuration space
${\cal C}$ with Markov transition rates from state $C$ to $C'$
$M_{C'\to C}:=M_{CC'}$. However, (\ref{eq:evolutionp_C}) \emph{can
not} in general be interpreted as a CTMC because the transition rates
$M_{CC'}$ in (\ref{eq:evolutionp_C}) are \emph{not necessarily positive}--see
(\ref{eq:EomsSpinhalf}) for an explicit example on a spin-$\frac{1}{2}$.
The mathematical definitions can be extended using negative probabilities
to force this interpretation but since negative probabilities have
no physical meaning--we cannot simulate an event occurring with negative
probability--this approach has limited practical uses. 

An alternative way forward is to think of the process as a CTMC with
\emph{negative transition rates} and keep the probabilities positive.
The key advantage of doing so is that there exists a systematic way
to map a CTMC with negative transition rates to a one with entirely
\emph{positive }transition rates at the cost of \emph{doubling }the
number of configurations. This procedure was proposed by V\"ollering
in \cite{Negativerates_Vollering} and we reintroduce it here now.%
{} We start from Eq.~(\ref{eq:evolutionp_C}) where the coefficients
$M_{CC'}$ can be negative. First, we define the \emph{entirely positive}
transition rates 
\begin{equation}
M_{CC'}^{\pm}:=\pm\theta\left(\pm M_{CC'}\right)M_{CC'},\label{eq:Mpm}
\end{equation}
where $\theta$ is the Heaviside function. Then, we double the configuration
space and denote \emph{particle }states with a $\bullet$ and \emph{antiparticle
}states with a $\circ$: ${\cal C}\to{\cal C}\times\left\{ \bullet,\circ\right\} .$
The probability $p_{C}$ of the original process is decomposed as
the difference 
\begin{equation}
p_{C}=p_{C}^{\bullet}-p_{C}^{\circ}.\label{probasum}
\end{equation}
One can readily check that Eq.~(\ref{eq:evolutionp_C}) can be obtained
from %

\begin{align}
d_{t}p_{C}^{\bullet/\circ}= & \sum_{C'\neq C}\left(M_{CC'}^{+}p_{C'}^{\bullet/\circ}-M_{C'C}^{+}p_{C}^{\bullet/\circ}+M_{CC'}^{-}p_{C'}^{\circ/\bullet}-M_{C'C}^{-}p_{C}^{\bullet/\circ}\right)+V_{C}p_{C}^{\bullet/\circ},\label{eq:probabilitesparticle/antiparticle}
\end{align}
with $V_{C}:=2\sum_{C'\neq C}M_{C'C}^{-}$. The previous
equations now constitute a well-defined Markov process with entirely
positive rates given by $M^{+}$ and $M^{-}$ and \emph{creation }rates
given by $V_{C}$. The \emph{transition rules} for the Markov process
corresponding to Eq.~(\ref{eq:probabilitesparticle/antiparticle})
are as follow: The elements of $M^{+}$ describe transition rates
for \emph{both }particles and anti-particles to move in the configuration
space. The element $M_{C,C'}^{-}$ is the probability rate for the
following event: a particle(antiparticle) in a configuration $C'$
moves to $C$ and \emph{is converted in its opposite }while two new
particles(antiparticles) are created on the original configuration
$C'$. These rules are represented schematically on Fig.~\ref{fig:Examples-of-transitions}
for the spin-$\frac{1}{2}$ case. Remark that the
$M^{+}$ transitions have the usual interpretation for CTMC where a
state with configuration $C'$ goes to $C$ while the $M^{-}$ terms can
be interpreted as a transition from $C'$ to $C$ carrying a \emph{negative
sign}.

Because of the creation term $V_{C}$, the number of (anti)particles
is not conserved anymore, i.e. replicas of the system are constantly
created. The state of the system for a given realization
of the process is now described by a collection of $2\times6^{N}$
numbers $n_{C}^{\bullet/\circ}$ indicating the occupation of configuration
$C$ in terms of particles and antiparticles. To enforce Eq.~(\ref{probasum}),
we impose that whenever a particle and antiparticle meet, they annihilate
each other. Hence, a single number $n_{C}:=n_{C}^{\bullet}-n_{C}^{\circ}\in\mathbb{Z}$
suffices to keep track of the state of the system. The configuration
space is then $\mathbb{\mathbb{Z}^{{\cal C}}}$ and a \emph{state}
$\eta\in\mathbb{Z}^{{\cal C}}$ is the collection of numbers $\left\{ n_{C}\right\} _{C\in{\cal C}}$.
The number of replicas ${\cal N}$ of the spin chain is $\sum_{C\in{\cal C}}\left|n_{C}\right|$.
Let $s_{C}:={\rm sign}\left(n_{C}\right)$ The update rules on the
replica space can be read from Eq.~(\ref{eq:probabilitesparticle/antiparticle})
to be: 
\begin{equation}
\eta\to\eta\pm s_{C'}\left(-\delta_{C'}+\delta_{C}\right)\text{ with probability \ensuremath{\left|n_{C'}\right|M_{CC'}^{\pm}dt}}.\label{eq:transition_rules}
\end{equation}
Let $f_{t}\left(\eta\right):\mathbb{Z}^{{\cal C}}\to\mathbb{C}$
be an observable on the replica space, the associated Kolmogorov backward
equation (the "Heisenberg" picture evolution in physicist language) is then
\begin{equation}
d_{t}f_{t}\left(\eta\right)=\sum_{\substack{C_{1}\neq C_{2},\\
s\in\left\{ +,-\right\} 
}
}\left|n_{C_{1}}\right|M_{C_{2}C_{1}}^{s}\left(f_{t}\left(\eta+ss_{C_{1}}\left(-\delta_{C_{1}}+\delta_{C_{2}}\right)\right)-f_{t}\left(\eta\right)\right). \label{eq:Kolmogorov_backward}
\end{equation}
The original probabilities are obtained by averaging
over different realizations of the process, i.e. 
\begin{equation}
p_{C}=\mathbb{E}\left[n_{C}\right].
\end{equation}
Remark that while $p_{C}$ is constrained to be positive, for a given
realization, $n_{C}$ can be `off-shell', i.e. may take any integer
value, positive or negative. The probability distribution that is always well-defined and normalized to $1$ is the probability of a given configuration $\eta$, $P_t(\eta)$ whose time evolution is given by the dual of \eqref{eq:Kolmogorov_backward}. Remark that this is very similar to what happens one goes from first to second quantization in quantum mechanics, except that one does not need to symmetrize or anti symmetrize the replica spaces in this case. Finally, we also note that the rules
(\ref{eq:transition_rules}) preserves the total relative number of
particles $\sum_{C}n_{C}$.

One conceptual advantage of having an explicit classical Markov process
is that it enables to simulate the system's evolution realization
by realization---something that is not feasible within the traditional
framework of quantum mechanics. A \emph{classical trajectory} is then
given by a realization of $\eta_{t}$, i.e. a set of occupation numbers
that fluctuate according to the transition rules. 

Remark that since particles and antiparticles are always created by
pairs, the difference between the total number of particles and antiparticles
is a conserved quantity for each realization, $\sum_{C\in{\cal {\cal C}}}n_{C}(t)={\rm Constant}.$

\section{State representation and expectation of observables}

We now explain how to represent quantum state and observables from
the classical CTMC point-of-view. 

A given density matrix $\rho$ encodes information about the probabilities
$p_{C}$ through the relationship $p_{C}=\frac{1}{m^{N}}\text{tr}\left(\rho\mathbb{P}_{C}\right)$.
In the classical model, this amounts to fix the average occupations
$\mathbb{E}[n_{C}]$. Since only the average is fixed, there exists
an infinite number of classical probability distributions for $\eta_{t}$
that average to $\rho$. When restricting to distributions involving
a single classical particle, this corresponds to randomly distributing
that particle across the configurations $C$ with probability $p_{C}$.

We now turn to observables. Let $\hat{O}$ be a quantum observable.
We restrict ourselves to strings of Pauli operators, $\hat{O}=\prod_{j\in A}\sigma_{j}^{\alpha_{j}}$
where $A$ is a set of $n$ different integers in $\left[1,N\right]$
and $\alpha_{j}\in\left\{ x,y,z\right\} $. $\hat{O}$ then admits
the decomposition: 

\begin{equation}
\hat{O}=\prod_{j\in A}\bigg(\sum_{s_{j}=\pm}s_{j}\mathbb{P}_{j}^{s_{j}^{\alpha}}\bigg)\prod_{k\in\bar{A}}\mathbb{I}_{k}.\label{eq:obs_decomposition}
\end{equation}
where $\bar{A}$ is the complement of $A$ in $\left[1,N\right]$
and we introduced the shorthand notation $s^{\beta}$ for $(s,\alpha)\in\left\{ +,-\right\} \times\left\{ x,y,z\right\} $.
The decomposition of $\hat{O}$ on the projectors $\mathbb{P}_{C}$
is not unique since any decomposition $\mathbb{I}=\sum_{s}\mu_{\alpha}\mathbb{P}^{s^{\alpha}}$
with $\sum_{\alpha}\mu_{\alpha}=1$ is a valid one. %
This gives us some ``gauge freedom'' to chose the value of the operator
on a configuration $C$. The symmetric decomposition corresponds to
the natural choice $\mu_{\alpha}=\frac{1}{m}$ $\forall\alpha$ so
that the value of the magnetization on a given site is independent
of the states on the other sites. However, nothing in principle prevents
us from making ``non-local'' choices. %

Once the gauge is fixed, we can attribute a \emph{classical value}
$O_{C}$ to each classical configuration $C\in{\cal C}$: $\hat{O}=\sum_{C\in{\cal C}}O_{C}\mathbb{P}_{C}$
and the quantum expectation value $\langle\hat{O}(t)\rangle:={\rm tr}(\rho_{t}\hat{O})$
is expressed as 

{} 
\begin{equation}
\langle\hat{O}(t)\rangle=m^{N}\sum_{C\in{\cal C}}O_{C}\mathbb{E}\left[n_{C}(t)\right].\label{eq:expobs}
\end{equation}

In order to make all the previous concepts clear, we now show how
they apply to the simplest case of a spin-$1/2$ rotation. %

\section{Spin-1/2 example}

Consider a spin-$\frac{1}{2}$ with $\left\{ \left|+^{\alpha}\right\rangle ,\left|-^{\alpha}\right\rangle \right\} $
as the $\sigma^{\alpha}$ eigenvectors with $\alpha\in\{x,y,z\}$.
We consider the evolution generated by the Hamiltonian
\begin{equation}
H=\frac{\sigma^{x}}{2}
\end{equation}
with initial state $\left|\psi(t=0)\right\rangle =\left|+^{z}\right\rangle $.
This simply describes the rotation of the spin around the $x$ axis
of the Bloch sphere with a period of $2\pi$. One way to encapsulate
this dynamics is through the Heisenberg equations of motion for the
expectations values of the projectors defined as $p_{\pm^{\alpha}}(t):=\frac{1}{2}{\rm tr}\left(\rho_{t}\mathbb{P}^{\pm^{\alpha}}\right)$,
$\mathbb{P}^{\pm^{\alpha}}:=\left|\pm^{\alpha}\right\rangle \left\langle \pm^{\alpha}\right|$.
As explained before, the factor $1/2$ is chosen so that the $p_{\pm^{\alpha}}(t)$
constitute a well-defined probability measure on the configuration
space with $4$ states ${\cal C}:=\left\{ +,-\right\} \times\left\{ y,z\right\} $
as they are all positive and sum to $1$. In terms of the notations
of the previous section, $m=2$ instead of $3$ as the motion is contained
in the $(y,z)$ plane. The Schr\"odinger equation (\ref{eq:Schrodinger})
translates into the system 

\begin{align}
d_{t}p_{+^{y}}=\frac{1}{2}\left(p_{-^{z}}-p_{+^{z}}\right), & \quad d_{t}p_{+^{z}}=\frac{1}{2}\left(p_{+^{y}}-p_{-^{y}}\right),\label{eq:EomsSpinhalf}
\end{align}
and the remaining quantities are obtained from the conservation of
probability $p_{+^{y/z}}+p_{-^{y/z}}=\frac{1}{2}$. %
The solution is 
\begin{equation}
p_{\pm^{y}}(t)=\frac{1}{4}\left(1\mp\sin t\right),\quad p_{\pm^{z}}(t)=\frac{1}{4}\left(1\pm\cos t\right).
\end{equation}
As explained before, one can not interpret (\ref{eq:EomsSpinhalf})
as a CTMC because of the signs of the coefficients. In order to do
so, we go to the doubled configuration space: %

\begin{equation}
\left\{ +,-\right\} \times\left\{ y,z\right\} \to\left\{ +,-\right\} \times\left\{ y,z\right\} \times\left\{ \bullet,\circ\right\} ,
\end{equation}
the transition matrices (\ref{eq:Mpm}) written in the basis $\left\{ +^{y},-^{y},+^{z},-^{z}\right\} $
are given by
\begin{equation}
M^{+}:=\frac{1}{2}\begin{bmatrix}0 & 0 & 0 & 1\\
0 & 0 & 1 & 0\\
1 & 0 & 0 & 0\\
0 & 1 & 0 & 0
\end{bmatrix},\quad M^{-}:=\frac{1}{2}\begin{bmatrix}0 & 0 & 1 & 0\\
0 & 0 & 0 & 1\\
0 & 1 & 0 & 0\\
1 & 0 & 0 & 0
\end{bmatrix}.
\end{equation}
This leads to the set of equations for the probability
\begin{align}
&d_{t}p_{+^{y}}^{\bullet/\circ}=\frac{1}{2}\left(p_{-^{z}}^{\bullet/\circ}+p_{+^{z}}^{\circ/\bullet}\right)-p_{+^{y}}^{\bullet/\circ}+p_{+^{y}}^{\bullet/\circ},\quad \\& d_{t}p_{+^{z}}^{\bullet/\circ}=\frac{1}{2}\left(p_{+^{y}}^{\bullet/\circ}+p_{-^{y}}^{\circ/\bullet}\right)-p_{+^{z}}^{\bullet/\circ}+p_{+^{z}}^{\bullet/\circ}. \nonumber\label{setofeq}
\end{align}
From these, one can now simulate the evolution of the system realization
by realization. The initial state $\left|+^{z}\right\rangle $ can
be obtained in the classical CTMC by drawing randomly the state of
a single particle among the configurations $\left\{ +^{z},+^{y},-^{y}\right\} $
with probabilities $\left\{ p_{+^{z}}^{\bullet}=\frac{1}{2},p_{+^{y}}^{\bullet}=p_{-^{y}}^{\bullet}=\frac{1}{4}\right\} $.
The particles are then moved, created and annihilated according to
the Markov transition rules defined previously which produces a classical
trajectory $\eta_{t}=\left\{ n_{C}(t)\right\} $. The expectation
value of e.g. $\sigma^{z}$ is then obtained from Eq.~(\ref{eq:expobs}):
$\left\langle \sigma^{z}(t)\right\rangle =2\mathbb{E}\left[n_{+^{z}}(t)-n_{-^{z}}(t)\right]$.

We show on Fig.~\ref{fig:Examples-of-transitions} the numerical
results for the spin-$\frac{1}{2}$. The different curves represent
different number of realizations of the process. We emphasize that,
even though it is clear now from our construction, it is remarkable
that we produced a classical stochastic process which gives rise to
oscillations of probabilities. This is typical of \emph{non-equilibrium
thermodynamics }as described by e.g. Lotka-Volterra prey-predator
equations \cite{LotkaOgpaper,VolterraOgpaper,Verhulstbook}.

\begin{figure}
\centering{}\includegraphics[width=1\textwidth]{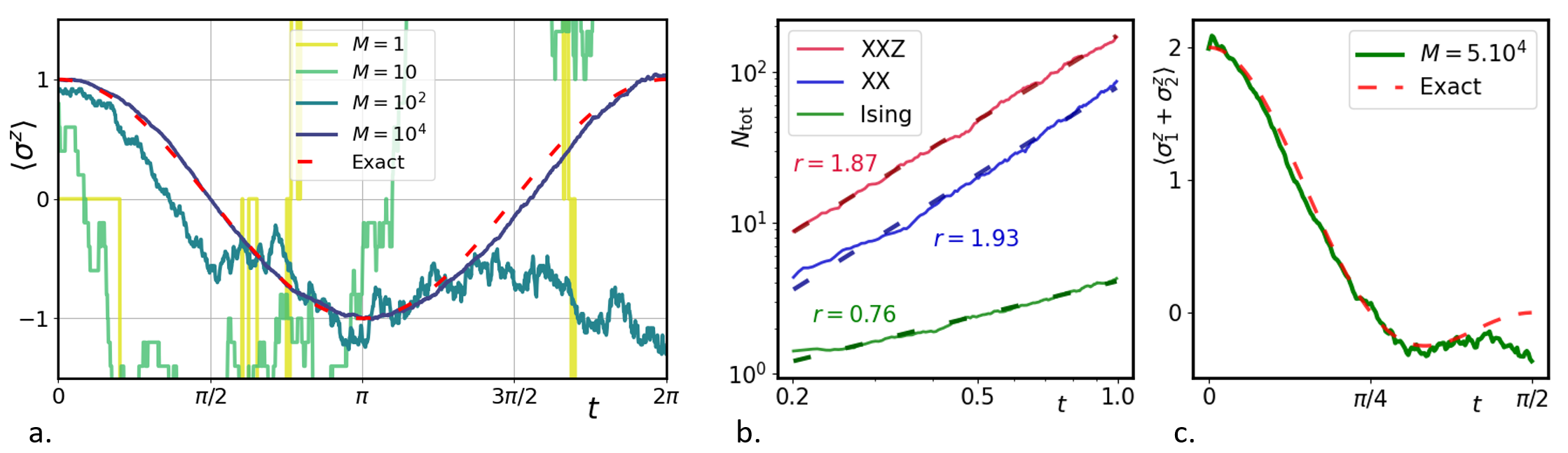}\caption{\textbf{a.} Evolution of $\left\langle \sigma^{z}(t)\right\rangle =2\mathbb{E}\left[n_{+^{z}}(t)-n_{-^{z}}(t)\right]$
as a function of time. $M$ indicates the number of realizations upon
which the classical process is averaged. The total number of particles
$N_{{\rm tot}}$ at $t=2\pi$ is $\sim9.8$ for this model. The red
dashed line is the exact solution $\left\langle \sigma^{z}(t)\right\rangle =\cos t$.
\label{fig:Examples-of-transitions-1}\textbf{b. }Averaged growth
of total number of particles $N_{{\rm tot}}$ for physical system
sizes $N=3$ with open boundaries as a function of time for the quantum
Ising, XX and XXZ models (\ref{Hamiltonians}). We average over 500,
50, 50 realizations for the quantum Ising, XX and XXZ models respectively
and take $\Delta=2/3$ for the XXZ. The initial state is taken as
the pure state with $\left|+^{z},-^{z},-^{z}\right\rangle $. We see
a sublinear growth at early times for the Ising model while the XX
and XXZ showcase an approximately quadratic growth. \textbf{c. }Time-evolution
of the half-magnetization $\left\langle \sigma_{1}^{z}+\sigma_{2}^{z}\right\rangle $
in the quantum Ising model for $N=4$ averaged over $M=5.10^{4}$
realizations. The initial state is taken as the quench $\left|+^{z},+^{z},-^{z},-^{z}\right\rangle $.
There is good agreement at early times while important fluctuations
due to the growth of $N_{{\rm tot}}$ renders the convergence at later
times more challenging. \label{fig:a.-Averaged-growth}}
\end{figure}

\section{Spin chains}

We now treat the important example of spin chains with pairwise interactions.
A generic Hamiltonian on a 1d lattice of $N$ sites is written as
\begin{equation}
H:=\sum_{j,k=1}^{N}\sum_{\alpha_{j},\alpha_{k}\in\left\{ x,y,z\right\} }t_{\alpha_{j},\alpha_{k}}^{j,k}\sigma_{j}^{\alpha_{j}}\sigma_{k}^{\alpha_{k}}.
\end{equation}
The transition matrix of the classical process for the pair of spins
$(j,k)$ is obtained by looking at the action of $H$ on the projectors
$\mathbb{P}_{j}^{s_{j}^{\beta}}\mathbb{P}_{k}^{s_{k}^{\beta}}$ .
From Pauli matrices algebra, one obtains the general formula: 
\begin{align}
i\left[\sigma_{j}^{\alpha_{j}}\sigma_{k}^{\alpha_{k}},\mathbb{P}_{j}^{s_{j}^{\beta}}\mathbb{P}_{k}^{s_{k}^{\beta}}\right]= & -\frac{1}{2}\sum_{s_{j}^{\beta'},s_{k}^{\beta'}}g{}_{s_{j}^{\beta'},s_{k}^{\beta'}}^{[\alpha_{j},\alpha_{k}]s_{j}^{\beta},s_{k}^{\beta}}\mathbb{P}_{j}^{s_{j}^{\beta'}}\mathbb{P}_{k}^{s_{k}^{\beta'}},\label{eq:formulaspins}
\end{align}
where we recall that $s^{\beta}$ is short for $(s,\beta)\in\left\{ +,-\right\} \times\left\{ x,y,z\right\} $
and $g^{[\alpha_{j},\alpha_{k}]}$ is defined as
\begin{align}
g_{s_{j}^{\beta'},s_{k}^{\beta'}}^{[\alpha_{j},\alpha_{k}]s_{j}^{\beta},s_{k}^{\beta}}:= & \varepsilon_{\alpha_{j},\beta_{j},\beta'_{j}}s_{j}s'_{j}\left(s'_{k}\delta_{\beta'_{k}}^{\alpha_{k}}+s_{k}\delta_{\beta'_{k}}^{\nu_{k}}\delta_{\alpha_{k}}^{\beta_{k}}\right)+\text{same term with }j\leftrightarrow k.
\end{align}
where $\varepsilon_{i,j,k}$ is the Levi-Civita symbol and the $\nu_{k}$s
can be arbitrarily chosen among the axis $\left\{ x,y,z\right\} $.
This comes from the fact that the identity $\mathbb{I}$ can be decomposed
as $\mathbb{I}=\mathbb{P}^{+^{\nu}}+\mathbb{P}^{-^{\nu}}$ for any
$\nu$. %
{} %
Once the $\nu_{k}$s are fixed, the Markov transition rates $M^{\pm}$
for the pair of spins $\left(j,k\right)$ from state $\left\{ s_{j}^{\beta'},s_{k}^{\beta'}\right\} $
to $\left\{ s_{j}^{\beta},s_{k}^{\beta}\right\} $ is given by $-\frac{1}{2}g_{s_{j}^{\beta'},s_{k}^{\beta'}}^{[\alpha_{j},\alpha_{k}]s_{j}^{\beta},s_{k}^{\beta}}$
and these rates encode the process for the whole chain. 

The whole process unfolds on the discrete space $\mathbb{Z}^{{\cal C}}$
whose size grows exponentially with physical system size $N$. Since the different realizations are independent, the standard error for the estimation of an observable $\langle \hat{O} \rangle$ is $\Delta\langle\hat{O}(t)\rangle\propto\frac{\sigma}{\sqrt{M}}$ where $\sigma$ is the standard deviation and $M$ the number of realizations which we average upon. For a local observable for instance, $\langle\hat{O}(t)\rangle$ is of order 1 while its value for a single realization is typically of order $N_{{\rm tot}}$, the
absolute total number of particles and antiparticles. Thus, a naive estimate of the error is
\begin{align}
\Delta\langle\hat{O}(t)\rangle\propto\frac{N_{{\rm tot}}}{\sqrt{M}}
\end{align}
Since the trajectory of a single particle is classically simulable,
the key challenge lies in the growth of $N_{{\rm tot}}$.
We show on Fig.~\ref{fig:a.-Averaged-growth}\textbf{b} examples
of the growth of particle number on the quantum-Ising, the XX and
the XXZ model with open boundaries, corresponding respectively to
\begin{align}
H_{{\rm Ising}}:=\sum_{j=1}^{N-1}\sigma_{j}^{x}\sigma_{j+1}^{x},\quad H_{{\rm XX}}:=H_{{\rm Ising}}+\sum_{j=1}^{N-1}\sigma_{j}^{y}\sigma_{j+1}^{y},\quad & H_{{\rm XXZ}}:=H_{{\rm XX}}+\Delta\sum_{j=1}^{N-1}\sigma_{j}^{z}\sigma_{j+1}^{z}.\label{Hamiltonians}
\end{align}
At early times, we observe that the growth of $N_{\text{tot}}$ in
the quantum Ising model is sublinear with respect to time, whereas
it exhibits a roughly quadratic growth in the XX and XXZ models. This
increase in $N_{\text{tot}}$ leads to significant oscillations across
the various contributions that must be summed over to compute an observable
of order 1. Consequently, a substantial number of realizations of
the process are necessary to achieve good convergence, which echoes
the challenge posed by the fermionic sign problem \cite{Signproblem_Scalettar1990,SignProblem_Troyer2005}. We also note that, while the growth in time of $N_{\rm tot}$ seems to be polynomial, a key question is the scaling with system size. Since the classical configuration space is exponentially large, we expect that the worst case generic scenario would be an exponential in system size prefactor in the growth in time. This question is left for a future thorough numerical exploration.

For the Ising model, we confirm that averaging over a sufficiently
large number of realizations accurately reproduces the quantum dynamics--see
Fig.~\ref{fig:a.-Averaged-growth}\textbf{c}. As expected, the convergence
deteriorates at later times due to the rising number of particles. 
\section{Conclusion}

In this paper, we propose a new way to describe the dynamics of quantum
spin chains in terms of purely \emph{classical }continuous time Markov
chains. Perhaps a contrario to the usual scenario, it is the quantum
dynamics that emerges from averaging over the classical one. Although
we restricted to spin chains, we expect generalizations to other types
of systems to be possible--if not straightforward. This offers a
fresh view on the dynamics of quantum system and will hopefully encourage
cross-fertilizations between the realms of classical stochastic processes
and quantum dynamics. We note in passing that our approach is reminiscent
of the many-world interpretation of quantum mechanics \cite{EverettManyworlds1957}
as the different classical copies can be thought of ``parallel universes''.
This also renders the theory \textit{nonlocal}, which
makes our formalism compatible with Bell's theorem \cite{Belltheorem}.

There are a lot of exciting questions that are raised by our study.
First and foremost, one should determine the usefulness of the stochastic
approach in solving the the many-body quantum problem. As explained
in the main text, our approach is not for now of much use due to the
increase of classical particle number during the time evolution. A promising direction stems from the fact that, due to the overcompleteness of the spin quantum basis, a given quantum theory can be "unraveled" into many classical processes on the replica space. This gauge freedom can then be exploited to reduce the computational complexity of the classical Markov process. These ideas will be developed and discussed further in \cite{LoiodeNardisJin_Q2C}.

The second important part of the quantum theory, apart from unitary
evolution, is projective measurement. One could of course re-implement
measurements the same way than in conventional quantum mechanics by
reconstructing the wave function from averaging over different realizations
and then applying Von Neumann rules for projective measurements. It
is also tempting however to implement measurement at the level of
the stochastic processes by describing it by a new non-reversible
CTMC where the transition rates are oriented towards the pointer states
corresponding to the observables measured. Such measurement protocoles
would reduce the total number of classical particles and could be
interesting for instance in the study of the dynamical interplay of
measurements and quantum dynamics \cite{BernardJinShpielberg_2018,caoEntanglementFermionChain2019,skinnerMeasurementInducedPhaseTransitions2019}.

\section*{Acknowledgements}
The author expresses his gratitude towards J.Brémont, J. De Nardis, C. Halati, L.Hruza, H. L\'oio, G. Morpurgo, L. Piroli,  A. Tilloy and X. Turkeshi for their feedback on the manuscript.  An early version of this project was presented at \textquotedblleft Les Gustins\textquotedblright{}
Summer School 2024. The author thanks all participants for their useful comments and criticisms. 

\bibliography{biblio.bib}


\end{document}